\documentclass[aps,prl,twocolumn,showpacs,superscriptaddress,groupedaddress]{revtex4-2}
\usepackage{graphicx}
\usepackage{bm}
\usepackage{amssymb}
\usepackage{amsmath}
\newcommand{\rev}[1]{\ifmmode\boldsymbol{#1}\else\textbf{#1}\fi}
\renewcommand{\Pr}{\mathit{Pr}}
\begin{document}
\title{Generalized Reynolds Analogy for Compressible Wall Turbulence: Unified Velocity--Temperature Relations from Mean to Fluctuating Field}
\author{You-sheng Zhang}
\affiliation{Institute of Applied Physics and Computational Mathematics, Beijing 100094, China}
\affiliation{Center for Applied Physics and Technology, HEDPS, College of Engineering, Peking University, Beijing 100871, China}
\affiliation{National Key Laboratory of Computational Physics, Beijing 100088, China}
\date{\today}
\begin{abstract}
The Reynolds analogy between velocity and temperature fields is a central problem in the statistical theory of compressible wall turbulence. The generalized Reynolds analogy (GRA) established by the author describes the mean velocity--temperature relation accurately, but a self-consistent closure for the fluctuating field has remained elusive. Here, the instantaneous similarity relation that the generalized total enthalpy (minus its wall value) is proportional to the local velocity is shown to close both fields at once. The mean field reproduces the GRA solution, while the fluctuating field yields the closed relation $T'_{rms}/u'_{rms} = f(Re_\tau,\Pr)\,\Pr_m^{-1/2}\,|\partial\bar{T}/\partial\bar{u}|$, where the prefactor $f$ is derived from the universal cascade dynamics of turbulence and the Obukhov--Corrsin theory, without adjustable parameters. The relation recovers the refined strong Reynolds analogy (RSRA) of Huang et al.~(2025), the most accurate benchmark to date, recasting it from an empirical relation into a consequence of universality principles and explaining the origin of its fitted prefactor $f=1.09$; and its ratio to boundary-layer and channel direct numerical simulation (DNS) data collapses onto unity across Mach numbers ($2.25$--$14$), Prandtl numbers ($0.025$--$1$) and wall thermal conditions, with an overall accuracy of about $5\%$.
\end{abstract}
\maketitle
The quantitative relation between velocity and temperature fields in compressible wall turbulence, known as the Reynolds analogy, underlies aerodynamic and thermal predictions in high-speed compressible flows. Since Morkovin\cite{morkovin1962effects} proposed the strong Reynolds analogy in 1962, it has succeeded under adiabatic conditions, but systematic deviations occur under strong wall heat transfer\cite{duan2011direct,pirozzoli2011turbulence}. Later models, such as those of Gaviglio\cite{gaviglio1987reynolds} and of Huang et al.\cite{huang1995compressible}, improved fluctuation predictions to varying degrees but remain largely empirical. Among recent models\cite{huang2025refined,cheng2024modified}, the refined strong Reynolds analogy (RSRA) of Huang et al.\cite{huang2025refined} is currently the most accurate empirical relation, yet it, too, lacks a closed derivation from underlying principles.

In 2014, the author and collaborators\cite{zhang2014generalized} established the GRA, a key breakthrough at the mean-field level. By introducing the generalized recovery factor $r_g$ and the effective turbulent Prandtl number $\overline{\Pr}_e$, the GRA provides analytical mean temperature--velocity relations validated by extensive DNS data\cite{zhang2014generalized,zhang2018direct} and incorporated into a recent textbook\cite{larsson2025turbulence}. A self-consistent closure for the fluctuating field within the GRA framework has, however, remained unattained\cite{huang2025refined,cheng2024modified}.

The central result of this paper is that both the mean-field GRA relation and the fluctuating-field statistics follow from the same instantaneous relation, namely, that the generalized total enthalpy (minus the wall value) is proportional to the local velocity. The mean field reproduces the GRA solution, while the fluctuating field yields Eq.~(\ref{eq:closed}), whose prefactor separates into a factor associated with the universal cascade process, which carries the complete dependence on the Reynolds number and the Prandtl number, and a structural similarity parameter $\Xi$, which is constant across the entire outer layer of each flow type. The mean and fluctuating fields thus cease to be independent problems and become manifestations of the same physical principle at different statistical levels.
\begin{figure}[!t]
\centering
\includegraphics[width=\columnwidth, trim=0cm 2cm 0cm 0cm, clip]{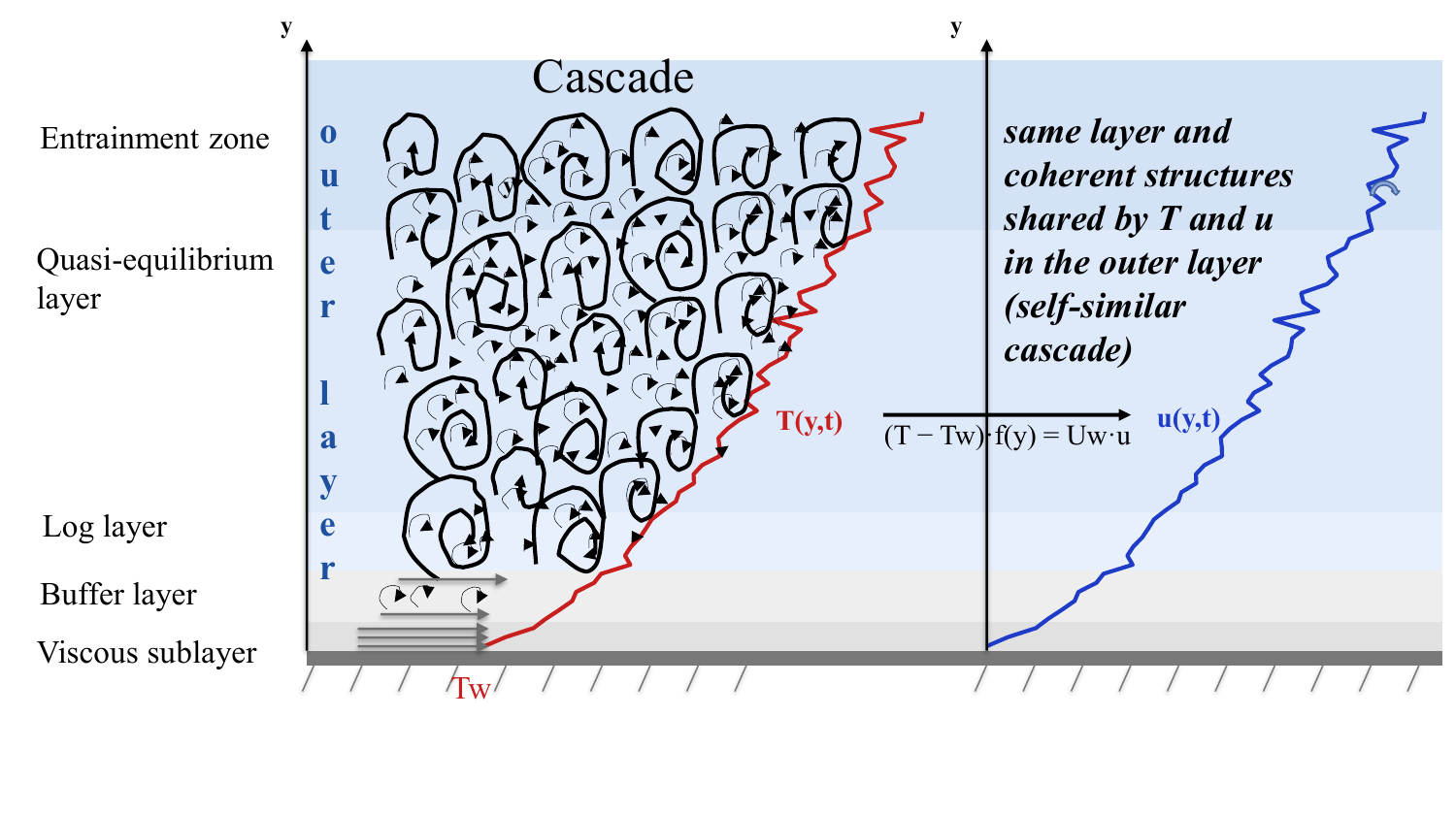}
\caption{Illustration of generalized similarity concept and cascade process of vortex in compressible wall turbulence.}
\label{fig:similarity}
\end{figure}

Figure~\ref{fig:similarity} illustrates the physical intuition of generalized similarity concept. In compressible wall turbulence, the instantaneous temperature $T$ and velocity $u$ profiles are not independent random fields; subtracting the wall temperature $T_w$ and stretching the ordinate via $f(y)$ maps the former pointwise onto the latter. This motivates a spatiotemporally dependent generalized recovery factor $r_g$ that renders the two profiles strictly proportional. We therefore define the generalized total enthalpy
\begin{equation}
H_g \equiv C_p T + r_g u^2/2,
\label{eq:H_g}
\end{equation}
where $C_p$ is the specific heat at constant pressure, and introduce the generalized similarity relation: at any instant and any location, the instantaneous generalized total enthalpy minus the wall enthalpy is proportional to the local velocity,
\begin{equation}
H_g - C_p T_w = U_w u,
\label{eq:similarity}
\end{equation}
with $U_w$ a global constant independent of the spatial coordinate $y$ and time $t$, while $u$, $T$, $r_g$ and $T_w$ all vary with $y$ and $t$. Since $r_g$ has sufficient degrees of freedom, such a proportionality coefficient can always be found, so that Eq.~(\ref{eq:similarity}) is a redefinition of $r_g$ rather than an approximation. This definitional, rather than assumptive, nature makes Eqs.~(\ref{eq:mean}) and (\ref{eq:fluct}) below exact identities; the physical input enters only at the closure developed thereafter.

Taking the ensemble average of Eq.~(\ref{eq:similarity}), with every instantaneous quantity $f$ decomposed into its Reynolds mean $\bar{f}$ and fluctuation $f'$, and retaining all fluctuation correlations, we obtain the exact mean-field equation
\begin{equation}
C_p\bar{T} + \bar{r}_g\bar{u}^2 /2+ \bar{M} - C_p\bar{T}_w = U_w\bar{u},
\label{eq:mean}
\end{equation}
where $ \bar{M} \equiv \bar{r}_g\overline{u'^2} /2+ \bar{u}\,\overline{r_g'u'} + \overline{r_g'u'^2}/2$ collects the second- and higher-order fluctuation correlations. Since the mean velocity and all fluctuations vanish at the wall, $\bar M$ and its derivative with respect to $\bar u$ vanish there, and differentiating Eq.~(\ref{eq:mean}) at the wall yields $U_w = C_p(\partial\bar{T}/\partial\bar{u})_w$.

Subtracting Eq.~(\ref{eq:mean}) from Eq.~(\ref{eq:similarity}) gives the exact fluctuating-field equation
\begin{equation}
T' + \phi' = A u',
\label{eq:fluct}
\end{equation}
where $A \equiv (U_w - \bar{r}_g\bar{u})/C_p = \partial\bar{T}/\partial\bar{u} +\epsilon_{r_g}+\epsilon_M$, with $\epsilon_{r_g}\equiv(\bar{u}^2/2C_p)\,\partial\bar{r}_g/\partial\bar{u}$ and $\epsilon_{M}\equiv(1/C_p)\,\partial\bar{M}/\partial\bar{u}$ obtained by differentiating Eq.~(\ref{eq:mean}) with respect to $\bar{u}$, and
$\phi' \equiv [\, \bar{r}_g(u'^2 - \overline{u'^2})/2 +r_g'\bar{u}^2 /2+ \bar{u}(r_g'u' - \overline{r_g'u'}) +(r_g'u'^2 - \overline{r_g'u'^2})/2\, ] /C_p- T_w'$.
Equation~(\ref{eq:fluct}) relates $T'$ linearly to $u'$ through the mean-field quantity $A$, while all nonlinear effects are concentrated into $\phi'$, which collects the velocity-fluctuation--recovery-factor couplings, third-order correlations, and wall-temperature fluctuations. This $\phi'$ is precisely the residual temperature fluctuation introduced, but not physically clarified, in the original GRA theory\cite{zhang2014generalized}. These two equations are not merely formal identities: within the quasi-one-dimensional approximation, the Reynolds-averaged Navier--Stokes equations admit them as self-consistent analytical solutions, as validated against DNS in the mean-field GRA theory\cite{zhang2014generalized}.

To derive a closed relation for the fluctuation intensity from Eq.~(\ref{eq:fluct}), we multiply Eq.~(\ref{eq:fluct}) by the wall-normal velocity fluctuation $v'$, by $u'$ and by $T'$ respectively and take Reynolds averages, obtaining $\overline{ v'T'}/(A\overline{ v'u'}) = 1-\eta_v$,\allowbreak $\overline{u'T'}/(A\overline{u'^2}) = 1-\eta_u$ and $\overline{T'^2}/(A\overline{u'T'}) = 1-\eta_T$, where $\eta_v \equiv\overline{v'\phi'}/(A\overline{v'u'})$, $\eta_u \equiv \overline{u'\phi'}/(A\overline{u'^2})$, $\eta_T \equiv \overline{T'\phi'}/(A\overline{u'T'})$. The key step is to form the ratio of the flux combination $\overline{v'T'}/(A\overline{v'u'})$ from the first relation to the variance combination $\overline{T'^2}/(A^2\overline{u'^2}) = (1-\eta_u)(1-\eta_T)$ from the product of the other two, isolating the effect of the residual field $\phi'$. This defines exactly the scalar $\Gamma(\phi',y) \equiv (1-\eta_u)(1-\eta_T)/(1-\eta_v)$, compressing the effect of $\phi'$ on all moments into a single quantity, giving $\overline{T'^2}/\overline{u'^2} = \Gamma\,A\,\overline{v'T'}/\overline{v'u'}$. With $\epsilon_{r_g}$ and $\epsilon_M$ negligible in the GRA mean-field framework, $A\approx \partial\bar T/\partial\bar u$, giving 
\begin{equation}
T'_{rms}/u'_{rms} = \sqrt{\Gamma}\,\Pr_{t}^{-1/2}\,|\partial\bar T/\partial\bar u|,
\label{eq:final}
\end{equation}
where $\Pr_t \equiv (\overline{ v'u'}/\overline{ v'T'})(\partial\bar{T}/\partial\bar{u})$ is turbulent Prandtl number. Eq.~(\ref{eq:final}) is a rigorous identity within the GRA framework, with $\Gamma$ an undetermined function of $\phi'$ and $y$, whose explicit form we now derive theoretically.

$\Gamma$ is not a mere formal scalar: rearranging Eq.~(\ref{eq:final}) shows it to be exactly the ratio of the two development time scales of Huang et al.\cite{huang2025refined}. For either field $f=u$ or $T$, define the development time $t_f \equiv f'^2/P_f$, with $f'^2$ the variance and $P_f$ its production rate ($P_{uu} \equiv -2\overline{u'v'}\,\partial\bar u/\partial y=2P_k$, with $P_k$ the kinetic-energy production, and $P_{TT} \equiv -2\overline{v'T'}\,\partial\bar T/\partial y$). Physically, $t_f$ is the time to sustain the local variance at the current production rate. One then has $\Gamma=t_T/t_u$. The problem is thereby converted into a physical one: what determines the ratio of the development times of temperature and velocity fluctuations? Since the two fields share the same cascade and differ only at the dissipation end, the answer must reduce to the arithmetic of one cascade with two cutoff scales.

To answer it, we first recall the structural picture of the boundary layer, sketched in Fig.~\ref{fig:similarity}. Along the wall normal the layer comprises the laminar viscous sublayer dominated by molecular viscosity, the buffer layer where laminar and turbulent motions alternate, and beyond them the outer layer dominated entirely by turbulent coherent structures\cite{pope2000turbulent}. Viewed from the structures themselves, the subregions of the outer layer are all filled with similar coherent structures. Viewed from the dynamical evolution of these structures, the evolution possesses overall similarity and universality: the energy of the turbulent fluctuations is injected at the large integral scale $L$, transferred essentially without loss through the inertial range, and dissipated by molecular processes at the cutoff scale\cite{kolmogorov1941local}. Both fluctuations are produced by their own mean gradients; at the moderate compressibility of present interest, temperature behaves essentially as a passive scalar, so the two fields share the same cascade path, the same eddies, the same integral scale $L$ and the same turnover time $\tau_L\equiv(L^2/\varepsilon_k)^{1/3}$ ($\varepsilon_k$ the kinetic-energy dissipation rate\cite{corrsin1951spectrum}). The only difference lies at the dissipation end: for fluids with $\Pr\equiv\nu/\alpha<1$ ($\nu$ the kinematic viscosity and $\alpha$ the thermal diffusivity), the velocity variance is cut off by viscosity at the Kolmogorov scale $\eta\equiv(\nu^3/\varepsilon_k)^{1/4}$, whereas the temperature variance is cut off earlier at the Obukhov--Corrsin scale $\eta_T=\eta\Pr^{-3/4}$\cite{corrsin1951spectrum}. These commonalities and this single difference form the basis of the following derivation and the physical root of its universality.

Within this picture, $\Gamma$ can be computed. For the velocity field, the inertial-range energy spectrum is $E_u(q)=C_k\varepsilon_k^{2/3}q^{-5/3}$\cite{sreenivasan1995universality}, where $q$ is the wavenumber and $C_k=1.62$ the Kolmogorov constant; the turbulent kinetic energy follows by integrating the spectrum from the injection scale $L$ to the cutoff scale $\eta$, giving $k=\int_{1/L}^{1/\eta}E_u(q)\,\mathrm{d}q=\frac{3}{2}C_k\varepsilon_k^{2/3}L^{2/3}(1-\chi)$, $\chi\equiv(\eta/L)^{2/3}$. For the temperature field, the Obukhov--Corrsin spectrum in the inertial--convective range\cite{obukhov1949structure,corrsin1951spectrum} is $E_T(q)=C_T\,\varepsilon_T\varepsilon_k^{-1/3}q^{-5/3}$, where $\varepsilon_T$ is the scalar dissipation rate and $C_T=0.76$, within the range reported for the Obukhov--Corrsin constant of the three-dimensional spectrum\cite{sreenivasan1996passive}; the temperature variance follows similarly by integration to its own cutoff scale $\eta_T$, giving $\overline{T'^2}=\int_{1/L}^{1/\eta_T}E_T(q)\,\mathrm{d}q=\frac{3}{2}C_T\varepsilon_T\varepsilon_k^{-1/3}L^{2/3}(1-\chi_T)$, $\chi_T\equiv(\eta_T/L)^{2/3}=\chi\Pr^{-1/2}$. With $f_k\equiv P_k/\varepsilon_k$ and $f_T\equiv P_{TT}/\varepsilon_T$ the production--dissipation ratios and $a_{uk}\equiv\overline{u'^2}/k$, the development times follow as $t_u=a_{uk}k/(2f_k\varepsilon_k)$ and $t_T=\overline{T'^2}/(f_T\varepsilon_T)$; substituting the two variance relations directly gives $\Gamma=C\,\Xi\,({1-\chi\Pr^{-1/2}})/({1-\chi})$,
where $C\equiv2C_T/C_k=0.94$ is a universal constant and $\Xi\equiv a_{kT}/a_{uk}$ with $a_{kT}\equiv f_k/f_T$.

The above relation shows that $\Gamma$ consists of two parts: the cascade factor $(1-\chi\Pr^{-1/2})/(1-\chi)$ and the structural factor $C\Xi$. For the former, $\chi=(\eta/L)^{2/3}$ remains to be determined. Attached-eddy theory and experiment\cite{marusic2013predictive} set the injection scale at $L=y_0$ in the logarithmic region, the characteristic location of the attached eddies; we take $y_0^+\equiv y_0/(\nu_w/u_\tau)=3\sqrt{Re_\tau}$, with $\nu_w/u_\tau$ the viscous length, $u_\tau$ the friction velocity, and $Re_\tau=u_\tau\delta/\nu_w$ ($\delta$ the layer thickness). With the logarithmic-region scalings evaluated there, $\varepsilon_k \sim u_\tau^3/(\kappa y_0)$ ($\kappa=0.41$) and $\eta/y_0=\kappa^{1/4}(y_0^+)^{-3/4}$, one obtains $\chi=(\eta/y_0)^{2/3}=\kappa^{1/6}(y_0^+)^{-1/2}\approx0.50\,Re_\tau^{-1/4}$. The cascade factor thus depends only on $\Pr$ and $Re_\tau$, tending to $1$ as $Re_\tau\to\infty$ or $\Pr\to1$.

For the latter part, besides the universal constant $C$, it contains the structural similarity parameter $\Xi$, which merits a detailed discussion. $\Xi$ is defined as the ratio of $a_{kT}$ and $a_{uk}$, the latter measuring the fraction of the turbulent kinetic energy carried by the streamwise fluctuation. In the author's earlier boundary-layer study, the author noted that although $\overline{u'^2}$ and $k$ each vary substantially along the wall normal, their ratio $a_{uk}\approx1.0$ remains constant from the near-wall region all the way to the outer edge $\delta$, independent of Mach number, wall temperature and Reynolds number\cite{zhang2012mach}; it was therefore identified as the Reynolds-stress structural parameter reflecting the boundary-layer organization. Inspired by this invariance, we extend the same idea to the production--dissipation ratios of the kinetic-energy and temperature-variance equations, $f_k$ and $f_T$, and find a similar conclusion: their ratio $a_{kT}\approx1.21$ remains constant beyond the viscous sublayer, and hence $\Xi_{BL}=1.21$. Furthermore, a similar invariance is found in channel flow: although the structural parameters $a_{uk}$ and $a_{kT}$ each vary along the wall normal, their ratio remains constant beyond the viscous sublayer, at the value $\Xi_{CH}\approx1.05$. We therefore term $\Xi$ the structural similarity parameter: it possesses an invariance one level higher than the quantities it comprises, the direct signature that momentum and heat transport in the outer layer are carried by the same coherent structures. It is a property of the flow rather than of the model: once the flow type is given, $\Xi$ is fixed a priori by its large-scale organization. Its constant values in the two systems differ by about $15\%$, a difference borne by the mean value of $a_{uk}$: in channels, very-large-scale motions fill the entire gap and pump energy predominantly into the streamwise component, raising the mean $a_{uk}$ to about $1.15$, whereas in boundary layers intermittent entrainment redistributes energy toward the transverse components and the mean $a_{uk}$ relaxes to about $1.00$. Correspondingly, $\sqrt{C\Xi}$ plays the role of the Corrsin mixing-length ratio $\ell_T/\ell_u$\cite{corrsin1951spectrum}: it equals about $1.0$ in channels, where heat and momentum are mixed over the same distance, and about $1.07$ in boundary layers, where the thermal mixing path is $7\%$ longer.

Eq.~(\ref{eq:final}) closes the core region; replacing $\Pr_t$ by the mixed Prandtl number $\Pr_m=(\nu+\nu_t)/(\alpha+\alpha_t)$\cite{walz1966boundary}, with $\nu_t=-\overline{u'v'}/(\partial\bar u/\partial y)$ and $\alpha_t=-\overline{v'T'}/(\partial\bar T/\partial y)$ the eddy viscosity and thermal diffusivity, extends it across the entire layer without patching, since $\Pr_m$ reduces to $\Pr$ at the wall and to $\Pr_t$ in the core. Near the wall, where DNS data place $T'_{rms}/u'_{rms}$ in the range $(1.0$--$1.3)\,|\partial\bar T/\partial\bar u|$, the prefactor approaches $\sqrt{\Gamma/\Pr_m}\approx1.25$ for $\Pr\approx0.7$ and $Re_\tau\sim10^3$.

Combining the above results, the final fluctuating relation reads
\begin{equation}
\frac{T'_{rms}}{u'_{rms}}=\sqrt{\frac{C\Xi}{\Pr_m}\frac{1-0.5Re_\tau^{-1/4}\Pr^{-1/2}}{1-0.5Re_\tau^{-1/4}} }\left|\frac{\partial\bar T}{\partial\bar u}\right|,
\label{eq:closed}
\end{equation}
with $\Xi=1.21$ for boundary layers and $\Xi=1.05$ for channels. For boundary layers at $\Pr\approx0.7$ and $Re_\tau\sim10^3$, Eq.~(\ref{eq:closed}) yields $\sqrt{\Gamma}=1.05$, very close to the RSRA fitted constant $1.09$, the most accurate empirical relation to date. The coefficient $1.09$ thereby receives a theoretical explanation: it is not a universal constant, but the joint value of the structural factor and the cascade factor in that parameter neighborhood, turning the RSRA into a consequence of the universal cascade organization and the Obukhov--Corrsin theory. The limit $\Gamma=1$ is realized when $\phi'=0$, i.e.\ perfect pointwise correlation $T'=Au'$, under which Eq.~(\ref{eq:closed}) reduces to the classical models of Gaviglio\cite{gaviglio1987reynolds} and Huang et al.\cite{huang1995compressible}; $\sqrt{\Gamma}$ therefore measures the strong but imperfect velocity--temperature correlation imposed by the large-scale structures.
\begin{figure}[t]
\centering
\includegraphics[width=0.96\columnwidth]{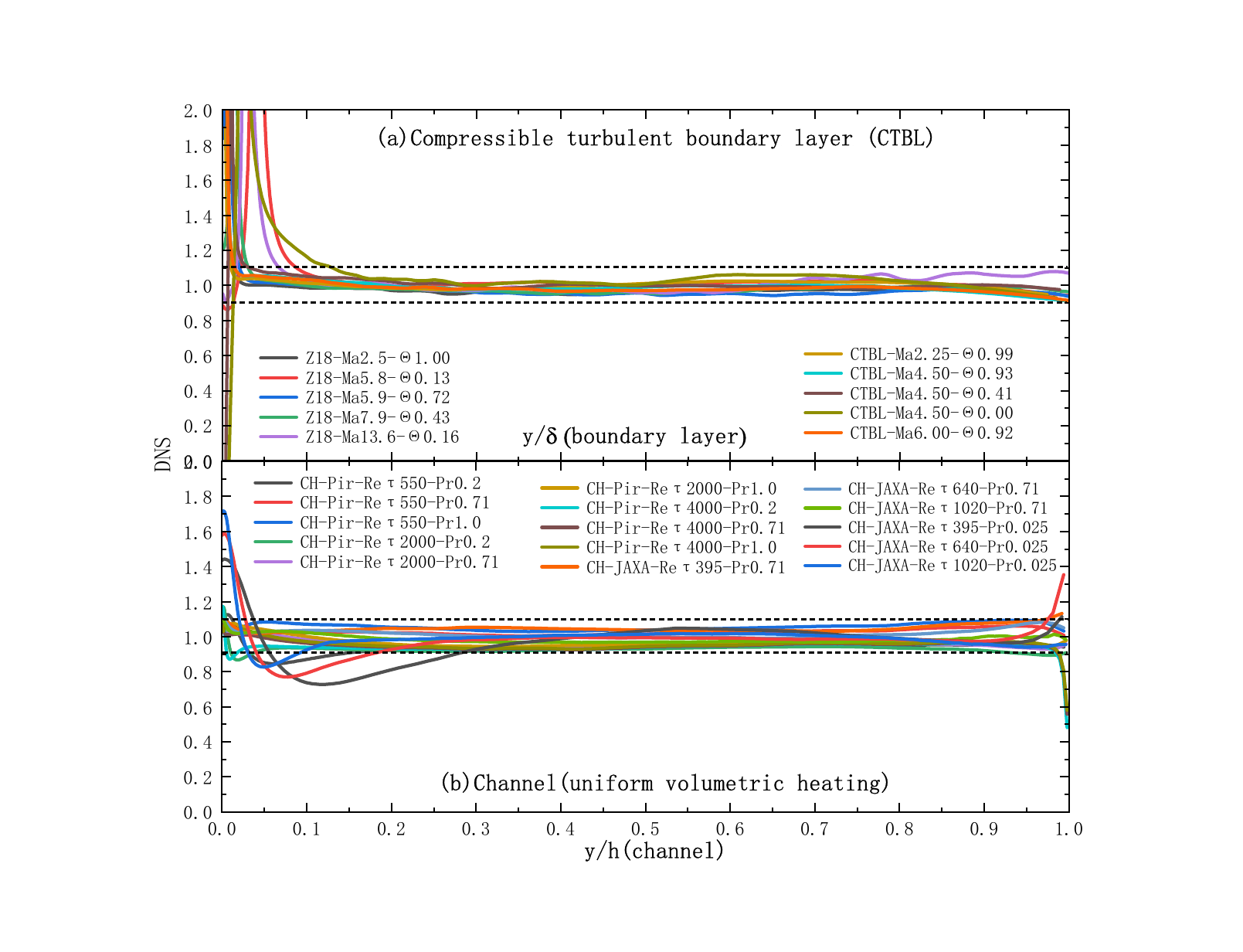}
\caption{Validation of the predicted fluctuation intensity ratio $T'_{rms}/u'_{rms}$ against DNS data. The ordinate is the ratio of the DNS value to the prediction of Eq.~(\ref{eq:closed}); the data of Zhang et al.\cite{zhang2018direct} and the channel data are evaluated with the turbulent Prandtl number $\Pr_t$, the mixed $\Pr_m$ being unavailable there; the two differ only below $y/\delta\approx0.2$, outside the validated range. (a) Boundary layers: the present compressible turbulent boundary layer data\cite{zhang2014generalized} ($\Pr=0.70$) and the data of Zhang et al.\cite{zhang2018direct} ($\Pr=0.71$); (b) channels with uniform volumetric heating: Abe et al.\cite{abe2009correlation} and Pirozzoli et al.\cite{pirozzoli2016passive}. The legends give the diabatic parameter $\Theta\equiv(T_w-T_\delta)/(T_r-T_\delta)$, with $T_\delta$ the edge and $T_r$ the recovery temperature.}
\label{fig:validation}
\end{figure}

Equation~(\ref{eq:closed}) is tested systematically against DNS data in Fig.~\ref{fig:validation}: adiabatic and diabatic boundary layers at $Ma=2.25$--$14$ and $Re_\tau=450$--$2350$, and channels with uniform volumetric heating at $\Pr=0.025$--$1.0$ and $Re_\tau=395$--$4088$, $25$ data sets in total. The DNS-to-theory ratio collapses onto unity throughout the outer layer, with root-mean-square deviation over $0.2\le y/\delta\le1$ of $3.2\%$ (boundary layers), $5.6\%$ (channels), and $5.1\%$ overall. The present validation does not include symmetric cold-wall channels, whose wall-heat-flux-induced contribution must first be subtracted\cite{zhu2025generalized}; this is left for future work. The derivation carries over unchanged to the Favre decomposition, and thus extends directly to strongly compressible flows. The mean field closes equally directly: with $\bar M$ neglected in the core region and the classical hypothesis $\overline{\Pr}_e=1$, Eq.~(\ref{eq:mean}) integrates analytically to the GRA mean temperature--velocity relation\cite{zhang2014generalized}.

This paper establishes a unified framework for the mean and fluctuating fields of compressible wall turbulence, building upon the GRA. The fluctuating-field prefactor separates into the structural similarity parameter $\Xi$, an invariant of the outer layer of each flow ($1.21$ in boundary layers and $1.05$ in channels) whose difference traces back to the streamwise energy share, and the cascade factor, depending only on $Re_\tau$ and $\Pr$, derived from the universal cascade organization and the Obukhov--Corrsin theory without adjustable parameters. The final relation recovers the RSRA as a local approximation, explains its coefficient $1.09$, predicts the suppressed fluctuation ratio of liquid-metal channels without adjustment, and is validated against $25$ DNS data sets across Mach numbers, Reynolds numbers, Prandtl numbers and wall thermal conditions, with an overall accuracy of about $5\%$. Both closures trace back to the same physical fact, that large-scale coherent structures are the common carriers of momentum and heat transport in the outer region. It provides a new basis for unified modeling of compressible turbulent boundary layers.

We thank Zhikang Huang, Wenfeng Zhou, Cheng Cheng and Yu Ming for discussions. Supported by NSFC under Grants No. 12588301 and 12532013.


\begin{thebibliography}{99}
\bibitem{morkovin1962effects}
M. V. Morkovin, in \emph{M\'ecanique de la Turbulence} (CNRS, Paris, 1962), pp. 367--380.
\bibitem{duan2011direct}
L. Duan and M. P. Martin, J. Fluid Mech. \textbf{684}, 25 (2011).
\bibitem{pirozzoli2011turbulence}
S. Pirozzoli and M. Bernardini, J. Fluid Mech. \textbf{688}, 120 (2011).
\bibitem{gaviglio1987reynolds}
J. Gaviglio, Int. J. Heat Mass Transfer \textbf{30}, 911 (1987).
\bibitem{huang1995compressible}
P. G. Huang, G. N. Coleman, and P. Bradshaw, J. Fluid Mech. \textbf{305}, 185 (1995).
\bibitem{huang2025refined}
Z. Huang, H. Su, Q. Guo, X. Yuan, X. Xiong, and Y. Zhou, J. Fluid Mech. \textbf{1019}, A13 (2025).
\bibitem{zhang2014generalized}
Y.-S. Zhang, W.-T. Bi, F. Hussain, and Z.-S. She, J. Fluid Mech. \textbf{739}, 392 (2014).
\bibitem{zhang2018direct}
C. Zhang, L. Duan, and M. M. Choudhari, AIAA J. \textbf{56}, 4297 (2018).
\bibitem{cheng2024modified}
C. Cheng and L. Fu, J. Fluid Mech. \textbf{999}, A20 (2024).
\bibitem{larsson2025turbulence}
J. Larsson and X. Zhong, Eds., \emph{Turbulence and Transition in Supersonic and Hypersonic Flows} (Academic Press, 2025).
\bibitem{zhang2012mach}
Y.-S. Zhang, W.-T. Bi, F. Hussain, X.-L. Li, and Z.-S. She, Phys. Rev. Lett. \textbf{109}, 054502 (2012).
\bibitem{kolmogorov1941local}
A. N. Kolmogorov, Dokl. Akad. Nauk SSSR \textbf{30}, 301 (1941).
\bibitem{obukhov1949structure}
A. M. Obukhov, Izv. Akad. Nauk SSSR, Ser. Geogr. Geofiz. \textbf{13}, 58 (1949).
\bibitem{corrsin1951spectrum}
S. Corrsin, J. Appl. Phys. \textbf{22}, 469 (1951).
\bibitem{walz1966boundary}
A. Walz, \emph{Str\"omungs- und Temperaturgrenzschichten} (Verlag G. Braun, 1966).
\bibitem{pope2000turbulent}
S. B. Pope, \emph{Turbulent Flows} (Cambridge, 2000).
\bibitem{marusic2013predictive}
I. Marusic, J. P. Monty, M. Hultmark, and A. J. Smits, J. Fluid Mech. \textbf{716}, R3 (2013).
\bibitem{sreenivasan1995universality}
K. R. Sreenivasan, Phys. Fluids \textbf{7}, 2778 (1995).
\bibitem{sreenivasan1996passive}
K. R. Sreenivasan and P. Kailasnath, Phys. Fluids \textbf{8}, 189 (1996).
\bibitem{abe2009correlation}
H. Abe, R. A. Antonia, and H. Kawamura, J. Fluid Mech. \textbf{627}, 1 (2009).
\bibitem{pirozzoli2016passive}
S. Pirozzoli, M. Bernardini, and P. Orlandi, J. Fluid Mech. \textbf{788}, 614 (2016).
\bibitem{zhu2025generalized}
Z. Zhu \emph{et al.}, J. Fluid Mech. \textbf{1012}, R2 (2025).
\end{thebibliography}
\end{document}